# The Age-Period-Cohort-Interaction Model for Describing and Investigating Inter-Cohort Deviations and Intra-Cohort Life-Course Dynamics*

Liying Luo

*Department of Sociology & Criminology*

*Population Research Institute*

*Pennsylvania State University*

James S. Hodges

*Division of Biostatistics, School of Public Health*

*University of Minnesota*

Version: April 2019

*An earlier version of this paper was presented at the 2014 meetings of the Population Association of America and the 2014 meetings of the American Sociological Association. We are grateful to Caren Arbeit, Julia Drew, Catherine Fitch, Zhi Li, Carolyn Liebler, Ian Ross Macmillan, Susan Mason, Robert M. O'Brien, and Rob Warren for providing helpful comments. This project has benefited from the support of the Population Research Institute, Pennsylvania State University and the Minnesota Population Center, University of Minnesota. Errors and omissions are the responsibility of the authors. Please direct correspondence to Liying Luo at liyingluo@psu.edu.

**Abstract**

Social scientists have frequently sought to understand the distinct effects of age, period, and cohort, but disaggregation of the three dimensions is difficult because cohort = period – age. We argue that this technical difficulty reflects a disconnection between how cohort effect is conceptualized and how it is modeled in the traditional age-period-cohort framework. We propose a new method, called the age-period-cohort-interaction (APC-I) model, that is qualitatively different from previous methods in that it represents Ryder's (1965) theoretical account about the conditions under which cohort differentiation may arise. This APC-I model does not require problematic statistical assumptions and the interpretation is straightforward. It quantifies inter-cohort deviations from the age and period main effects and also permits hypothesis testing about intra-cohort life-course dynamics. We demonstrate how this new model can be used to examine age, period, and cohort patterns in women's labor force participation.

**Introduction**

Social scientists are often concerned about trends. For example, have Americans become politically more liberal or conservative in recent decades? Have death rates in the US declined across birth cohorts? Do Americans' vocabularies decrease as they become older and has this changed over the past decades? Answering questions like these requires analysts to consider simultaneously the roles of three distinct dimensions of time: *age* (how old people are at the time of interview), *period* (the year in which they are interviewed), and *cohort* (in these examples, the year in which they were born). Meanwhile, in a society in which individual biographies are shaped by social characteristics such as race, gender, and socioeconomic status, people who differ with these respects can show distinct age, period, and cohort patterns in social, demographic, and economic outcomes. Therefore, investigating age, period, and cohort patterns can provide new insights about how aging, social changes, and population processes interact with social institutions such as schools and families to (re)produce inequality.

To separate the independent effects of age, period, and cohort, Mason et al. (1973) proposed an age-period-cohort (APC) accounting model. Unfortunately, no unique estimates exist for this APC accounting model because the value for one of the three variables is completely determined by the other two: cohort = period – age. That is, researchers have sought to understand three dimensions of time, yet one dimension is an exact function of the other two dimensions.

The problem in the classical APC model is widely viewed as a statistical challenge and various technical solutions have been proposed to circumvent it. Alongside methodologists questioning the validity of the assumptions imposed by those techniques (Glenn 2005; Luo 2013; Luo and Hodges 2016; O'Brien 2011; Rodgers 1982), we argue that the supposedly technical difficulty in identifying the traditional APC model is in fact theoretical in nature. That is, the type of cohort effect that is estimated in the classical APC accounting model differ from the concept of cohort effect that interests social scientists. This implies that even if effect estimates could be obtained by imposing a constraint on the APC accounting model, they may not be meaningful or interpretable; the standard interpretation is not sensible because the third variable (e.g., cohort) cannot vary holding the other two constant (e.g., age and period).

Drawing on the literature of sociology, demography, and biostatistics, we propose a new age-period-cohort model, called the age-period-cohort-interaction (APC-I) model. The APC-I model is conceptually different from previous APC methods in that it defines cohort effects as the differential effects of social change (i.e., period effects) depending on one's age, whereas by including cohort as an additive predictor, previous methods developed under the traditional APC accounting framework arbitrarily assume that cohort effects can occur even when social events affect individuals of different ages in a uniform way.

Methodologically, the APC-I model has three advantages. First, it is fully identified and flexible enough to include other important social and demographic predictors. Second, it gives meaningful estimates that reflect theoretical ideas about age, period, and cohort effects. Third, it allows researchers to examine life-course dynamics within cohorts, an important type of cohort-related variation that is not accessible using the APC accounting model.

Our research also contributes to the emerging discussions about linear and nonlinear effects in APC analysis. Statisticians have noted that whereas linear effects of age, period, and cohort are inseparable, nonlinear (e.g., quadratic, cubic, or quartic) effects are identifiable (Holford 1983). Some researchers have subsequently recommended focusing on the nonlinear components of cohort effects (Chauvel and Schröder 2015; O'Brien, Hudson, and Stockard 2008)

However, we show that under the classical APC accounting framework, nonlinear cohort effects may still be problematic as they may include nonlinear age and period effects. This also clarifies the fundamental difference between the APC-I model and previous methods that impose a constraint on the linear components of the effects.

This paper proceeds as follows. We begin by contrasting the concept of cohort effect that interests social scientists and the cohort effects, linear and nonlinear, that are estimated in the classical APC accounting model; the two are not the same. Next, we introduce the new APC-I model, provide theoretical and methodological rationales for it, describe how it is specified, and explain how inter- and intra-cohort effects can be estimated and tested. We then use an example of female labor force participation to demonstrate how to apply the model in empirical research. We conclude by discussing the application of the APC-I model in particular situations, its connections with other APC models, and its potentials and limitations.

**Cohort Theories and the APC Accounting Model: Disparities between conceptualization and operationalization**

*Cohort Analysis and the Age-Period-Cohort Framework*

Researchers in many disciplines are interested in how outcomes vary across time in a society in which individual biographies are shaped by social events and historic shifts. For example, demographers have studied social conditions that affect temporal trends in divorce rates in the US (Kennedy and Ruggles 2014). For another example, sociologists of religion have attempted to test the theory of secularization, which refers to the decline of religion in modern societies (Chaves 1989; Firebaugh and Harley 1991). Until the 1970s, research on temporal processes was dominated by an age-period paradigm, a paradigm that only considers changes with age and time periods. *Age* is arguably one of the most important factors in social science research: a wide range of research has documented that many social, demographic, economic, and health outcomes change as one gets older (Borella, Ghisletta, and de Ribaupierre 2011; Cole 1979; Elder 1975; Lynch et al. 2004). At the same time, social and historical changes, captured as a package by *period effects*, can affect individual outcomes and attributes including political views, vocabulary knowledge, and health conditions (Peng 1987; Smith 1990; Wilson and Gove 1999; Winship and Harding 2008).

Demographers and sociologists have challenged this age-period paradigm, arguing that this type of research ignores an important dimension of temporal processes: *cohort*. A cohort refers to a group of individuals who experience a significant event like birth, marriage, or graduation at the same time. Cohort is a key concept and useful analytical tool because cohort differences reflect the formative effects of exposure to social events during critical ages, which act persistently over time (Ryder 1965). Social science literature has demonstrated the importance of cohort; omitting cohort in analyzing temporal trends may lead to spurious conclusions about age and period patterns.

Earlier investigation of cohort differences involved a simple, descriptive comparison of cohort-specific age profiles (Greenberg, Wright, and Sheps 1950; Schaie 1965). Although this simple approach can provide important clues about observed cohort differences, period effects are not removed and thus may be confounded with cohort effects. Therefore, answering questions about temporal processes of many variables requires analysts to simultaneously consider the distinct effects of age, period, and cohort. To formally estimate and infer the independent effects of age, period, and cohort, Mason et al. (1973) specified an analysis of variance (ANOVA) model that they titled the age-period-cohort (APC) accounting model[1]:

$$g(E(Y_{ij})) = \mu + \alpha_i + \beta_j + \gamma_k, \qquad (1)$$

for age groups $i = 1,2,\ldots,a$, periods $j = 1,2,\ldots,p$, and cohorts $k = 1,2,\ldots,(a+p-1)$, where $\sum_{i=1}^{a} \alpha_i = \sum_{j=1}^{p} \beta_j = \sum_{k=1}^{a+p-1} \gamma_k = 0$. $E(Y_{ij})$ denotes the expected value of the outcome $Y$ for the $i$th age group in the $j$th time period; $g$ is the "link function"; $\alpha_i$ denotes the mean difference from the global mean $\mu$ associated with the $i$th age category; $\beta_j$ denotes the mean difference from $\mu$ associated with the $j$th period; $\gamma_k$ denotes the mean difference from $\mu$ associated with membership in the $k$th cohort. The usual ANOVA constraint applies where the sum of coefficients for each effect is set to zero. Unfortunately, the APC accounting model has methodological and theoretical limitations, which the next two sections discuss.

*Methodological Critique: What the APC accounting model estimates*

It is well known that the APC accounting model (1) is not identified because of the linear dependency between age, period, and cohort; see Fienberg and Mason (1979); Fosse and Winship (forthcoming); Luo et al. (2016) for detailed discussions. Briefly, the design matrix of model (1) has rank one less than full, so an infinite number of solutions (estimates) for the parameters $\alpha, \beta,$ and $\gamma$ fit any data equally well. That is, the data cannot distinguish different estimation results, so a constraint—in addition to the usual reference group or sum-to-zero constraint—must be imposed in order to choose one set of estimates.

---

[1] This is called an "accounting" model because age, period, and cohort are just indicators of underlying forces such as education expansion and economic recessions that actually cause the phenomena that one observes.

This identification problem is inherent in any APC model that attempts to separate independent, additive effects of age, period, and cohort and thus cannot be solved by changing the model setup (e.g., using random effects for period and cohort as in Yang and Land 2008) or by variable manipulation (e.g., using unequal interval width for age, period, and cohort groups as in Robertson and Boyle 1986 and Sarma et al. 2012). The identification problem is well recognized and its consequences have been discussed extensively (Fienberg and Mason 1979; Fosse and Winship Forthcoming; Kupper et al. 1983, 1985; Luo et al. 2016; Luo and Hodges 2016). In essence, internal information derived from the data cannot help either because the problem is circular: researchers do the analysis to learn precisely the kind of information needed to justify any such constraint. Therefore, scholars have emphasized that the choice of the constraint must be based on theoretical grounds or external information (Fienberg 2013; Glenn 2005; Luo 2013; O'Brien 2013).

Such theoretical information, however, may be difficult to obtain. Moreover, even when a constraint can be justified on theoretical grounds, the effect estimated obtained from the APC accounting model (1) can be difficult to interpret for two reasons. First, the standard interpretation of multiple regression coefficients—the pure association between one predictor and the outcome holding other predictors fixed—is not sensible because there can be no variation in, say, age when period and cohort are held fixed, and analogously for period and cohort.

Second, suppose that each of the age, period, and cohort effect has linear and quadratic trends[2], then model (1) can be written as

$$Y = \beta_0 + \beta_1 a + \beta_2 a^2 + \beta_3 p + \beta_4 p^2 + \beta_5 c + \beta_6 c^2 + \varepsilon, \tag{2}$$

where $Y$ is the outcome, $\beta_0$ denotes the grand mean, and $\beta_1, \beta_2, \ldots, \beta_6$ denote the coefficients of the linear and quadratic age, period, and cohort terms. Because $cohort = period - age$, replacing cohort terms with age and period results in

$$Y = \beta_0 + \beta_1 a + \beta_2 a^2 + \beta_3 p + \beta_4 p^2 + \beta_5 (p-a) + \beta_6 (p-a)^2 + \varepsilon. \tag{3}$$

---

[2] The use of a continuous term to index cohort membership may seem odd to some analysts; we use this strategy only to demonstrate the implication of the age-period-cohort linear dependency for estimating and interpreting cohort effects in the classical APC accounting model.

Simple algebra then gives[3]

$$Y_{ij} = \beta_0 + \beta_1 a + \beta_2 a^2 + \beta_3 p + \beta_4 p^2 + \beta_5 (p-a) + \beta_6 (a^2 + p^2 - 2 \cdot a \cdot p)^2 + \varepsilon_{ij}. \tag{4}$$

Eq. (4) shows that the "cohort effects" that APC model (1) attempts to estimate in fact involve linear age and period effects, quadratic age and period effects, and most crucially an age-by-period interaction. In fact, any one of the three effects can be expressed as a combination of linear and nonlinear effects of the other two.

Eq. (4) is revealing because it shows that even when researchers *can* determine a set of estimates (i.e., a constraint on $\beta$) based on theoretical grounds, the resulting estimates for cohort effects are a combination of linear and nonlinear age and period effects and their interaction. This is unfortunate because the APC accounting model is designed to simultaneously isolate the supposedly independent effects of age, period, and cohort, but apparently it has not achieved this goal.

---

[3] Alternatively, Eq. (4) can be written as $Y_{ij} = \beta_0 + (\beta_1 - \beta_5)a + (\beta_2 + \beta_6)a^2 + (\beta_3 + \beta_5)p + (\beta_4 + \beta_6)p^2 - \beta_6(2 \cdot a \cdot p)^2 + \varepsilon_{ij}$. Clearly, while this alternative expression of Eq. (4) is identified, $\beta_1, \beta_3$, and $\beta_6$ are not; that is, infinitely many possible sets of $(\beta_1, \beta_3, \beta_6)$ give identical estimated coefficients in the alternative expression of Eq. (4).

The identification problem discussed above has been characterized as methodological in nature (see, e.g., (Glenn 1976, 1989, 2005; Mason et al. 1973)). With that focus, inadequate attention has been given to the theoretical problem that creates the methodological problem in the APC accounting model. In the following subsection, we argue that the APC accounting model fails not so much because of the identification problem but because it makes a conceptual error by assuming that there *are* independent, additive age, period, and cohort effects in phenomena of interest.

*Theoretical Critique: How a cohort effect is defined*

Whereas a cohort refers to a group of individuals, a cohort *effect* manifests because of the unique experiences of social events and historical shifts at one's formative ages. Most eloquently, Norman Ryder (1965) offered a theoretical vision about how cohort effects manifest in his seminal work on cohort analysis:

> The aggregate by which the society counterbalances attrition is the birth cohort, those persons born in the same time interval and aging together. Each new cohort makes fresh contact with the contemporary social heritage and carries the impress of the encounter through life. … The new cohorts provide the opportunity for social change to occur. They do not cause change; they permit it. If change does occur, it differentiates cohorts from one another, and the comparison of their careers becomes a way to study change. <u>The minimal basis for expecting interdependency between intercohort differentiation and social change is that change has variant import for persons of unlike age</u> [emphasis added], and that the consequences of change persist in the subsequent behavior of these individuals and thus of their cohorts. (1965: 844)

He further elaborated three basic notions on which cohort analysis rests:

> persons of age *a* in time *t* are those who were age *a*-1 in time *t*-1; transformations of the social world modify people of different ages in different ways; the effects of these transformations are persistent. In this way a cohort meaning is implanted in the age-time specification. (1965: 861)

According to this conceptualization, a cohort effect is *defined* as the interaction between age and period effects, where "interaction" is understood in the sense used by statisticians. A social or historical transformation that has uniform consequences for people of all ages can have no cohort effect; likewise, an age-related process that works the same way in all time periods also cannot have a cohort effect. Conceptually, this differs from thinking about cohort as having an independent, additive effect net of period and age effects. While researchers have sought (at least implicitly) to isolate the independent effect of cohort among people who are equivalent with respect to age and period, in the new APC model that we introduce below, we *conceptualize a cohort effect as the degree to which age and period effects are moderated by one another*.

What does this concept of cohort effect mean for describing and explaining temporal trends in APC analysis? Instead of assuming that period effects do not exist or that cohort has independent, additive effects net of age and period effects, we argue that a researcher should begin by explicitly describing the degree to which age effects vary across time periods or equivalently, the extent to which period effects vary across age groups. If the effects of period are the same across age groups or equivalently, if the effects of age are the same across periods, she must look for explanations for trends of interest that do not rely on cohort processes. If, however, such moderating effects are present, then the researcher must seek explanations that are consistent with this empirical pattern. It seems very likely, for example, that temporal changes in church attendance have occurred differently in different age groups; older people's church-going activity is probably less amenable to change, and church attendance of younger people may be declining. If so, cohort should be a useful variable for explaining trends in church attendance—but this is the case only if the effects of period vary by age and or vice versa, which is the definition of a statistical interaction between age and period.

*Intra-Cohort Life-Course Dynamics: Cumulative, diminishing, or constant?*

Another theoretical limitation of the APC accounting model and its variants is that they ignore life course dynamics as a cohort ages, i.e., they assume cohort effects remain fixed or constant across the life course (Hobcraft, Menken, and Preston 1982). That is, previous research using these models not only assumes cohort has an independent, additive effect net of age and period, but also that this cohort effect does not change for individuals across their life course. However, such constant cohort effects may not be plausible. For example, being a young adult when the civil rights movement swept through America may have a lasting effect on individual's political views, but it is not necessary to assume that this effect persists without change into later life for that birth cohort. That is, intra-cohort life-course dynamics, an important type of cohort-related variation, are ignored by the APC accounting model.

Fortunately, under Ryder's (1965) conceptualization of cohort effect and in the new model introduced below, the assumption of time-constant cohort effects can be relaxed so that competing theoretical ideas about intra-cohort life-course variation can be examined in empirical studies. For example, the "cumulative advantage/disadvantage" theory (Dannefer 1987, 2003; DiPrete and Eirich 2006; Ferraro and Kelley-Moore 2003) resonates with the "Matthew Effect" (Merton 1968) and the saying "the rich get richer; and the poor get poorer" (Entwisle, Alexander, and Olson 2001). It posits that the initial (dis)advantages of people with different capacities, resources, and structural locations are incremental or cumulative over the life course so that gaps in social, economic, health, or cognitive outcomes between the advantaged and disadvantaged tend to widen over the life course. For example, the protective effects of higher education on overall health may be multiplicative over the life course as highly educated people can avail themselves of more resources and thus have better opportunities than the less educated to adopt and maintain a healthier life style and behaviors (Cheng 2016; O'Rand 1996; Pampel and Hunter 2012)

In contrast, the age-as-leveler hypothesis, represented by "the survivor effect" in mortality research (Hobcraft et al. 1982), argues that a harsh environment in early life may eliminate vulnerable individuals of a cohort so that cohort would show, for example, higher death rates and worse general health in young ages but lower mortality rates and better health when they are old (Beckett 2000; Dupre 2007; Ferraro and Farmer 1996).

Such within-cohort life-course dynamics also have important implications for understanding the ways in which period shocks may occur and materialize.[4] For example, the Great Recession in the late 2000s may negatively affect the overall health of the 1980s birth cohort through the mechanisms of stress, reduced income, and diminished employment opportunity in their early careers, but the negative influences may not manifest until they reach older ages. Such lagged effects of time periods may not be detected if one focuses on average differences between cohorts as in the traditional APC accounting framework. In contrast, investigating variations over a cohort's life course may provide valuable clues about when a period shock begins to differentiate cohort experiences.

---

[4] We thank an anonymous reviewer for offering this insight.

Because the classical APC accounting model focuses on differences between cohorts and assumes such differences are fixed across the life course, researchers using this model give up the opportunity to test intra-cohort life-course hypotheses including cumulative (dis)advantage or age-as-leveler theories. We show below how to use the APC-I model to test this important but neglected type of cohort-related variation.

**Toward a Paradigm Shift: A new model**

The preceding discussion of the methodological limitations of the APC accounting model and associated estimation techniques is not to deny the theoretical importance or explanatory power of the concept of a cohort. The point, rather, is that any search for an ultimate statistical solution under the APC accounting framework, attempting to estimate cohort effects independent of age and period effects, is a "futile" and "unholy" quest (Glenn 1976:900 and (Fienberg 2013):1981, respectively). All forms of the APC accounting model, including the intrinsic estimator (Yang et al. 2004) and hierarchical APC method (Yang and Land 2008), have serious limitations because they conceive of cohort effects in a way that departs from the theoretical account of Ryder (1965) and because they assume little life-course variation. To address these problems, researchers must move beyond the accounting framework and precipitate a paradigm shift (Kuhn 1996).

We propose a new model that is conceptually and methodologically different from other APC methods, an APC model that explicitly estimates and tests cohort effects as age-by-period interactions. Each of the hypotheses about intra-cohort life-course dynamics, "constant effects", "cumulative (dis)advantage", or "age-as-leveler", corresponds to a specific structure of the age-by-period interaction. This new APC model is closely tied to theoretical ideas about cohort effects, it is fully identified, and it is flexible enough to test various hypotheses about life-course changes within cohorts. Thus we believe it is a step towards a paradigm shift in APC research. We first describe the model specification and rationale and then suggest appropriate estimation and testing techniques. We then show how the new model can be used to test theoretical ideas about inter- and intra-cohort changes using the example of women's labor force participation.

*Model Specification*

The APC accounting model (1) implies that an independent, additive cohort effect can be present even when the effects of period apply equivalently to all age groups. However, as discussed above, sociological and demographic theories imply that cohorts are not differentiable unless period effects differ between age groups. Informed by this theoretical insight, we propose a new model, called the age-period-cohort-interaction (APC-I) model, that treats cohort effects as a specific form of the age-by-period interactions. The general form of this model can be written as

$$g(E(Y_{ij})) = \mu + \alpha_i + \beta_j + \alpha\beta_{ij(k)}, \tag{5}$$

where $g, Y_{ij}, \mu, \alpha_i$ and $\beta_j$ are defined as in model (1) and $\alpha\beta_{ij(k)}$ denotes the interaction of the $i$th age group and $j$th period group, corresponding to the effect of the $k$th cohort. Note that the effect of one cohort includes multiple age-by-period interaction terms $\alpha\beta_{ij(k)}$ that lie on the same diagonal in a table with ages in rows and periods in columns.

Model (5) differs from model (1) in the way cohort effects are modeled; here, cohort effects are considered as a specific form of the age-by-period interaction (we return to this point in the next paragraph). In statistics, the interaction between two variables describes the differential effects of one variable depending on the level of the other variable (Scheffé 1999). In APC research, this means that if the temporal patterns of interest can be attributed to cohorts, significant age-by-period interactions should be present. When cohort membership is not associated with the outcome—that is, when the effects of historical or social shifts (period effects) are uniform across age categories—then an age-by-period interaction is not present.

The view that cohort effects can be quantified as the age-by-period interaction has not gone unnoticed in the APC literature. For example, Clogg (1982) noted that cohort effects "are special kinds of A-P interaction" (p. 467). Also, Holford (1983) argued that "[a] model which assumes that … there is an additive effect due to age, period and cohort is in itself arbitrary. We might instead have considered interactions, but in fact if we look at interactions among any two factors, the third factor [e.g., cohort, in the APC accounting model] spans a subspace of that interaction space" (p. 322).

Technically, it is indisputable that the effects of any choice of third variable from among age, period, and cohort can be expressed as the interaction between the other two variables. In this sense, the APC-I model appears to privilege age and period effects and "discriminate" against cohort effects by including age and period main effects and reducing cohort effects to the interaction of the other two. The theoretical reason for our choice to include age main effects is that researchers are usually interested in a general age pattern that many individuals follow as they get older. Period main effects are used to represent the impacts of social changes that everyone in the society is exposed to. The decision to explicitly quantify cohort effects as a specific form of age-by-period interaction is informed by the literature on how cohort effects are conceptualized in relation to age and period effects. Empirically, as the analyses of women's labor force participation in the following section will show, the size of the cohort effects, characterized as the age-by-period interaction, is not necessarily smaller—in fact may be larger—than some of the main effects. In many cases and many disciplines, interaction terms often have no substantive meanings and thus are difficult to interpret. APC analysis represents an opportunity to model and interpret the age-by-period interaction terms in a meaningful and sensible way.

In fact, the conceptual motivation of the APC accounting model is similar to the APC-I model. As Fienberg and Mason (1985) were well aware, "the inclusion of a set of cohort effects in this kind of model [the APC accounting model] is a way to get a simple and parsimonious description of age by period interactions" (1985: 71), although this simple and parsimonious description comes at the price of the identification problem and cohort effects that are constant over the life course. To illustrate, suppose we have a normally-distributed outcome $Y$ with five age categories and five periods. Table 1's top panel represents the expected value of the outcome $E(Y_{ij})$ in each cell in terms of unknown effects $\alpha_i$ and $\beta_j$ in the classical APC accounting model (1) and includes $5 - 1 = 4$ independently-varying estimates for the five cohort effects. Table 1's bottom panel represents the expected value in each cell in terms of the parameters $\alpha_i, \beta_j$ and $\alpha\beta_{ij(k)}$ in the APC-I model (5). The latter includes $(5 - 1) \cdot (5 - 1) = 16$ independently-varying estimates for the $5 \cdot 5 = 25$ age-by-period categories, where the remaining $25 - 16 = 9$ quantities are computed using the usual ANOVA constraints.

[Table 1 About Here]

Consider, for example, the 5$^{th}$ cohort, in the diagonal that runs from the upper-left to the lower-right cell. The effect of belonging to that cohort in the top panel of Table 1, $\gamma_5$, corresponds to five elements in the age-by-period interaction, $\alpha\beta_{11(5)}, \alpha\beta_{22(5)}, \alpha\beta_{33(5)}, \alpha\beta_{44(5)}$, and $\alpha\beta_{55(5)}$, in the bottom panel. The latter five age-by-period interaction terms are unrestricted in model (5), meaning that they can take on any values (subject to summing to zero down columns and across rows). In model (1), these five age-by-period interaction terms are replaced by a single parameter $\gamma_5$, conforming to a particular theory about changes over the life course within a cohort. Therefore, the APC accounting model, at least conceptually, may be viewed as a special case of the APC-I model that attempts to recover a special type of cohort effect by replacing the $(a \cdot p)$ age-by-period interactions with $(a + p - 1)$ cohort categories. However, this parsimony is costly: its price is the model's identifiability and ability to investigate within-cohort dynamics. The developers of the APC accounting model recognized this limitation (Fienberg and Mason 1985: 70, 84), but unfortunately many APC researchers have taken the accounting model as the final word and thus focused on solving the identification problem.

In the "Estimation and Testing" subsection that begins below we describe statistical procedures that we propose to estimate and test cohort effects characterized as age-by-period interactions. This section is fairly technical, so one can skip it on a first reading. In outline, we introduce a three-step procedure for estimating and testing the overall age-by-period interaction, the set of interaction terms that correspond to each cohort, and inter- and intra-cohort differences.

*Estimation and Testing*

The concept of characterizing cohort effects as the age-by-period interaction described in the preceding section could be implemented in more than one way. In this section, we describe one way to estimate and test these effects and the interpretations that it allows. However, other ways are possible, and we consider them in "Discussion".

With cohort effects represented as the age-by-period interaction, testing hypotheses about inter- and intra-cohort variation is equivalent to examining a specific form of—that is, specific patterns and structures in—the diagonal cells of an age-by-period cross-classification after removing age and period effects. Specifically, in the APC-I model, variation between cohorts can be summarized using each cohort's average deviation from the age and period main effects.

Variation *within* cohorts can be investigated by examining specific contrasts of the age-by-period interactions that correspond to each cohort. The APC accounting model's assumption of a constant $\gamma_5$ is equivalent to testing a specific pattern in $\alpha\beta_{11(5)}, \alpha\beta_{22(5)}, \alpha\beta_{33(5)}, \alpha\beta_{44(5)}$, and $\alpha\beta_{55(5)}$. We describe below a three-step procedure for investigating age and period effects and inter-cohort deviations and intra-cohort dynamics. The next section demonstrates this procedure with an empirical example. Exemplary R code for the tests in Steps 1 through 3 are available upon request.

Step 1. A global deviance test: Are there variations in the outcome of interest associated with cohort membership that cannot be explained by age and period effects? To answer this question, fit model (5), which includes age main effects, period main effects, and their interaction. Then test the variation attributable to the age-by-period interaction, with $(a-1)(p-1)$ degrees of freedom. A significant global test statistic indicates that some kind of cohort effects *may* be present. Note that a significant global test does not characterize cohort effects, nor is it a sufficient condition for the existence of cohort effects. For example, the interaction might appear significant because the data deviate from pure age and period main effects but do so in a haphazard manner, i.e., one that has no reasonable interpretation as a cohort pattern.

In theory, a significant test is a necessary condition for cohort effects: A non-significant result suggests that the age-by-period interaction does not explain much variation in the outcome, so the reduced model with only age and period main effects fits the data as well as the model with the full interaction. In other words, a non-significant test suggests that the effects of social events are not differential for individuals of different ages so that there is no evidence that cohort membership matters for the outcome of interest. In this case, there may be no need to do the tests in Steps 2 or 3, which examine cohort patterns. Meanwhile, we recommend a visual inspection of the age-by-period interactions to facilitate interpreting age and period main effects (see Remark 2 at the end of this section for more details).

Step 2. Deviation magnitude tests: Does membership in a specific cohort matter, after accounting for age and period main effects? We can address this question using a deviance test about the set of the age-by-period interactions that corresponds to each cohort. This deviance test examines the magnitude of cohort-specific deviations from age and period main effects; that is, whether that cohort's group of age-by-period interactions, taken together, explains a significant proportion of variation in the outcome. If the deviance test rejects the null hypothesis in such a test, one may conclude that membership in that cohort has effects. However, these deviance tests do not allow researchers to distinguish to what extent or in what ways cohorts differ *from each other* in the outcome of interest. Steps 3.1 and 3.2 include two $t$ tests for characterizing between-cohort differences and within-cohort life-course dynamics.

Step 3.1. Average deviation tests. For each cohort that significantly deviates from age and period main effects based on Step 2, compute the average of the age-by-period interaction terms contained in that cohort and use a $t$ test to examine the average of that cohort-specific deviation. These averages and associated $t$ tests can be used to assess differences between cohorts in terms of their deviation from the age and period main effects.

Step 3.2. Life-course dynamics tests. For each cohort that deviates from the pattern defined by the age and period main effects, conduct a $t$ test of the linear (and quadratic if desirable) orthogonal polynomial contrast of the cohort's age-by-period interaction terms to investigate whether the average (dis)advantages of members of that cohort accumulate, remain stable, or diminish in their life course.

Table 2 provides a guideline about how to use the results of Steps 3.1 and 3.2 to evaluate three hypotheses about within-cohort dynamics, "constant effects", "cumulative advantage/disadvantage", and "age-as-leveler" hypotheses. Specifically, the data can be considered to support the cumulative (dis)advantage hypothesis when a given cohort's average (Step 3.1) and linear trend (Step 3.2) have the same sign, as shown in Table 2's upper-left and lower-right cells. When a cohort's average deviation and linear trend have opposite signs, as in Table 2's upper-right and lower-left cells, this supports the "age-as-leveler" hypothesis: a cohort's initial (dis)advantage diminishes as that cohort ages. When a cohort's average deviation is not statistically significant but its linear trend is significant, this also favors the leveling theory. If the linear trend is not significant but the average deviation is significant, then the constant effects hypothesis—the hypothesis implicit in the APC accounting model—seems plausible *for that cohort*. If neither the average deviation nor the linear trend is significant, it could mean that there is no clear pattern in cohort variation, and the significant deviation magnitude test is a result of some kind of deviation that does not conform to any theoretical idea of cohort effects.

[Table 2 About Here]

Four remarks about the three-step procedure: First, the idea of using these test statistics in APC analysis is not new. For example, Clayton and Schifflers 4/19/19 11:35:00 AM recommended using deviance or likelihood-ratio tests to choose among an age-only model, an age-period model, an age-cohort model, or a full age-period-cohort model. Also, Yang (2008) suggested comparing these models, though arguing that certain test outcomes justify using a constrained approach like the intrinsic estimator. However, the purpose of the global deviance test proposed here is neither model selection nor verifying technical constraints on the unknown parameters. Rather, because we consider cohort effects as the interaction between age and period, the global deviance test of the age-by-period interaction in model (5) serves as an explicit measure of and necessary condition for cohort effects.

Second, in the presence of a significant age-by-period interaction, researchers should use caution in interpreting estimated age and period main effects. In general, there are two types of interactions: a quantitative interaction, in which the trend of the outcome in age, say, has the same direction for all periods but periods differ in the strength of the trend; and a qualitative or cross-over interaction, in which the trends in age have different directions depending on period. It may be difficult to interpret main effects in a meaningful way when cross-over or qualitative interactions are present. However, one can still interpret main effects in the presence of qualitative interactions, as an average trend (Aiken and West 1991; Jaccard and Turrisi 2003). In the APC-I method, age and period main effects are interpretable with quantitative age-by-period interactions. With qualitative age-by-period interactions, we recommend data visualization to assess the magnitude of the differences in age patterns for each period and vice versa. At a minimum, researchers should refrain from reifying age and period main effect estimates when the effect of period depends qualitatively on age to such an extent that a general age pattern or period pattern, applied to the whole population, is not meaningful. The interpretation decision should be made case by case. An alternative sensible analytical strategy in the presence of qualitative age-by-period interactions is a comparison of cohort-specific age-graded trajectories because in this case, each cohort has a distinct aging process and is subject to the impacts of social change in a way differing from other cohorts.

Third, researchers should be careful about interpreting cohort effects when two or fewer age-by-period cells make up the cohort's diagonal in the age-period classification table, usually for the youngest or the oldest cohorts, because it may be misleading to treat a trend determined by so few data points (e.g., two age-by-period cells) as a good indicator of the general trend for that cohort across its life course. The more age-by-period cells observed for a cohort, the more informative the estimates are for understanding life course changes within that cohort.

The fourth remark concerns the implications of coding schemes (e.g., dummy coding as opposed to effect/sum-to-zero coding) for interpreting the age-by-period interaction. Although for some APC estimators, cell means estimates and conclusions change with coding (Grotenhuis et al. 2016; Luo et al. 2016), in any identified model including the APC-I model, the estimates for different coding schemes are equivalent and the cell means estimates are the same for all coding schemes. That is, although the numerical values of effect estimates and the interpretations naturally differ between coding schemes (e.g., dummy coding versus effect coding), the two sets of interaction estimates can be transformed to be equivalent so that the cell means in an age-period combination after considering the main effects are the same and conclusions do not depend on coding schemes. In other words, although the two interaction estimates necessarily differ in numerical values, the difference does not arise from an identification problem, but from a shift in meaning that these quantities represent. For example, whereas under the dummy coding an interaction represents a directional difference with a particular reference group, in effect coding an interaction represents the deviation from the cell mean implied by the age and period main effects. For the APC-I model, we recommend the effect coding for the purpose of easy interpretation and to be consistent with the recommendation of coding schemes in the presence of interactions (see Aiken and West 1991; Jaccard and Turrisi 2003), although this choice is not necessary.

**Temporal Patterns in Black and White Women's Labor Force Participation 1990-2017: An empirical application**

In this section we demonstrate how to use the APC-I model and the estimation and testing strategies described in the preceding section to examine age, period, and inter-cohort deviations and intra-cohort changes in labor force participation among black and white women between 1990 and 2017. Here, we do not attempt to provide a full assessment of social and demographic factors responsible for trends in their participation rates; the main objective of this exercise is to demonstrate how our method can be used in practically important and sociologically interesting questions.

*Background*

Temporal trends in women's labor force participation (LFP) have been the subject of much research (see, e.g., (Connelly 1992; Farkas 1977; Hollister and Smith 2014; Macunovich 2012; Treas 1987). While female LFP in the US continued to grow until the 1990s, as Fig. 1 shows, the growth appeared to reach a plateau in the 2000s and began to decline afterwards. This leveling-off or decline has sparked policy discussions and scholarly debates about the causes and implications of such trends. For example, scholars have attributed the observed leveling off to period-specific factors including labor demand (Erceg and Levin 2014), the economic shocks of the Great Recession (Boushey 2005; Hoffman 2009), social welfare and disability insurance (Duggan and Imberman 2009), and gender role attitudes (Fortin 2015).

[Fig. 1 about here]

However, this phenomenon is unlikely a pure period process. For example, because women usually begin to exit the labor force in their 50s, we should expect a decline in LFP if the female population is aging, that is, the proportion of women aged 50 or older has increased. That is, the recent trends may reflect structural change in the age composition of the female population (Aaronson et al. 2014). The cohort replacement process may also contribute to the stalled LFP rates through older cohorts with higher participation rates exiting and younger cohorts with lower participation rates entering the labor force (Lee 2014). Meanwhile, important social and demographic shifts such as educational attainment, number of children, attitudes, and beliefs are more likely to be a cohort-specific than period-specific process as these forces and changes mostly affect individuals at certain ages (Balleer, Gomez-Salvador, and Turunen 2014; Farré and Vella 2013; Fernández 2013; Goldin 2006).

Decomposing these trends into age-, period-, and cohort-related variation can therefore help illuminate the nature of the temporal trends and provide important clues about demographic, social, and economic factors that give rise to these temporal variations. In previous studies that examined age, period, and cohort effects in women's LFP, researchers were forced to make difficult-to-verify assumptions about one or two of these effects (see, e.g., Balleer et al. 2014; Clogg 1982; Euwals, Knoef, and Vuuren 2011; Lee 2014). In this demonstration, we use the APC-I model, which does not require such assumptions, to summarize the temporal variation in women's LFP in all three dimensions. In particular, we show how the APC-I model could be used to investigate intra-cohort life-course variation, examining whether advantages or disadvantages of a cohort in LFP are constant, cumulative, or diminishing over the life course *over* and *above* the general age and period pattern. Such exercises are timely as LFP scholars have begun to discuss how cohorts differ in their life-cycle shifts (Goldin and Katz 2018; Goldin and Mitchell 2017).

Specifically, we attempt to answer two sets of research questions:

1. For each race group (black and white), (a) to what extent does LFP vary as a function of the three dimensions of time, i.e., age, period, and cohort, from 1990 to 2017? (b) Over and above age and period general patterns, which cohorts have especially high and especially low LFP rates relative to their age and period? (c) Are cohort differences in LFP constant, cumulative, or reduced over each cohort's life course? That is, are some cohorts consistently more likely to engage in the labor force relative to their age and the time periods that they experienced? Or do cohorts demonstrate distinct life course patterns of LFP as they age?

2. For each race group, in what ways have changes in educational attainment and number of children affected those patterns? That is, would LFP rates be higher, stagnant or lower had education and number of children not changed in the United States?

*Data*

We use data from the 1990 through 2017 Current Population Survey (CPS) March Supplement (as disseminated by IPUMS-CPS). The CPS is a monthly survey conducted by the Census Bureau and the Bureau of Labor Statistics. A battery of questions on demographics and labor force participation is fielded every month. The focal outcome is labor force participation. Every year since 1962, CPS has asked respondents whether they participated in the labor force during the week prior to the interview. Being in the labor force (coded 1) means the respondents "were at work; held a job but were temporarily absent from work due to factors like vacation or illness; were seeking work; or were temporarily laid off from a job during the reference period" (Flood et al. 2017). The respondents were otherwise out of the labor force (coded 0). Age and year of interview are ascertained in every survey.

We selected working-age women 20 to 64 years old[5] in the 1990 to 2017 CPS samples and excluded respondents with missing values for LFP, age, survey year, gender, race[6], educational attainment, or number of children, giving a sample of 1,213,497 records[7] for whites and 181,064 for blacks. We constructed nine age groups (20-24, 25-29, … 55-59, and 60-64), six periods (1990-1994, 1995-1999, 2000-2004, …, 2010-2014, and 2015-2017), and thus 14 birth cohorts (1930, 1935, …, 1990, 1995)[8]. Table 3 presents descriptive statistics for the LFP and for race, educational attainment, number of children, and the three time-related predictors (age, period, and cohort).

---

[5] The LFP of the youngest (16-19 years old) and oldest (65 and older) age groups have much variation and irregularities, so following Clogg (1982), we omitted these age groups.

[6] While race categories in the CPS changed in 1988, 2003, and 2013, the black and white categories are comparable across the full 1990-2017 time period.

[7] A person who appears in one March CPS will also appear in an adjacent March CPS. Therefore, the sample size in this research refers to the number of individual records, not respondents.

[8] In a table of five-year age groups and five-year periods, a birth cohort is defined by diagonals of the age-period cross-classification table and extends over a nine-year interval. For example, the observations in 1995 through 1999 for people in the 30 to 34 age group describe the birth cohort of 1961 to 1969. Conventionally, each cohort is identified by its mid or central birth year (e.g., Mason and Winsborough 1973; O'Brien 2011). We follow this practice so, for example, the 1945 cohort refers to the group of people born between 1941 and 1949. When so defined, birth cohorts overlap with adjacent cohorts. This overlap is usually ignored in statistical modeling (Kupper et al. 1985).

[Table 3 About Here]

*Results*

To describe the temporal variation in female LFP, we fit separate APC-I models[9] that include age and period main effects and their interaction without covariates for each race group using the CPS data[10]. Following Step 1, we conducted a formal deviance test to detect possible cohort effects characterized as age-by-period interactions. The *F* statistics for the interactions are 47.773 for whites and 3.197 for blacks, and both are statistically significant ($p<0.001$). This means that for both ethnicity groups, the model that includes the age-by-period interaction fits the data better than the model with age and period main effects only. As we explained, for theoretical and methodological reasons, we consider cohort effects as specific kinds of age-by-period interactions. We thus conclude that there *may* be cohort effects on LFP rates; that is, an APC-I model should be more helpful than a reduced model that omits the interactions to accurately describe temporal changes in white and black women's LFP.

We further investigated the implications of those age-by-period interactions for interpreting age and period main effects by graphically plotting LFP probabilities over the ages for each time period in the left panel and the probabilities across periods for each age group in the right panel based on the weighted logistic APC-I model results reported in Table 4. An age-by-period interaction—i.e., some kind of cohort effect—would be visible in these plots as one or more lines with a shape that is different from other lines. Fig. 2 shows that for both whites and blacks, although the shapes of the age trajectories (left panel) change relatively little across the six time intervals, the period trajectories in LFP (right panel) differ between the younger and the older age groups. Specifically, for those aged 55 and older, LFP continues to grow until the 2000s, in contrast to an earlier plateau appearing in the 1990s among the younger groups. This visual inspection complements and supports the technical conclusion based on $F$ tests that cohort effects, characterized by certain kinds of age-by-period interaction effects, may be an important source of variation. We note that because of the somewhat different directions of the period trends for the older and the younger among whites, one may find it difficult to interpret an overall period trend. We discuss alternative modeling and interpretation approaches in "Discussion and Conclusion". For now, because the period patterns for different age groups are not completely opposite, we proceed to describe a general trend for age and period, i.e., the age main effects and period main effects.

---

[9] We used the effect or sum-to-zero coding in all the models. Alternatively, one could use the reference-group coding (e.g., omitting the first group). As noted earlier, analysis results using different coding schemes are equivalent.

[10] Because the CPS uses a complex sampling strategy, all analyses used the weighting variable "WTSUPP" provided by IPUMS-CPS.

[Table 4 about here]

[Figure 2 about here]

Models 1a and 2a in Table 4 report and Fig. 3a plots age and period main effects estimates without including any covariates in the APC-I model; these models describe "raw" age and period patterns in LFP. LFP appears to increase with age through the 40s and decline thereafter for both race groups, although the decline begins at an older age for whites (late 40s) than for blacks (late 30s). On average, blacks have higher LFP rates in the 30s and early 40s, whereas the LFP among whites slightly exceeds blacks' when they are age 50 or older. The estimated period effects in Models 1a and 2a in Table 4 and Fig. 3a suggest that on average, black and white women's LFP leveled off between 1990 and 2017. In general, the size of the period effects is smaller than that of the age effects.

[Fig. 3 about here]

To answer the question about whether and how cohort membership affects LFP, we examined the age-by-period interaction terms—cohort deviations from the LFP trajectory defined by age and period main effects—in the APC-I model. Table 5 presents estimated age-by-period interaction terms in APC-I Model 1a for whites and Model 2a for blacks, with rows defined by age groups and columns by time periods. As explained in Table 1, the interaction terms on each diagonal $\alpha\beta_{ij(k)}$ correspond to the estimated deviations of cohort $k$ from the main effects of age group $i$ and period $j$. For example, the age-by-period interaction terms that lie on the upper-left to lower-right diagonal for white women aged 20-24 in 1990-94 though 45-49 in 2015-17 (i.e., 0.152, 0.061, …, -0.099) correspond to the deviations of the 1970 birth cohort from the mean log odds LFP implied by the age and period main effects. Such deviations cannot be explained by the age and period main effects and can thus be uniquely attributed to their experiences as a cohort. All but a few of the interactions for whites in the top panel of Table 5 are substantially and statistically significant, indicating that many white women cohorts' LFP rates deviate significantly from their general age and period pattern. In contrast, only a few black cohorts appear to diverge as much as whites as shown in the bottom panel of Table 5. This is consistent with the global $F$ test in Step 1 where the $F$ statistics are 47.773 for whites and 3.197 for blacks.

[Table 5 about here]

We then followed Steps 2 and 3 to summarize and formally test these interactions; that is, to identify cohorts that had especially high or low LFP and whether their (dis)advantages changed over the life course, as summarized in Table 2. We conducted deviation magnitude tests (Step 2) on the set of interaction terms corresponding to each cohort. Table 6 reports the test results for models without covariates (Models 1a, 2a), with education covariates (Models 1b, 2b), and with number of children (Models 1c, 2c). As described above, the deviation magnitude test for a given cohort compares the fit of a simple model with only main age and period effects versus a more saturated model with all the main effects and *o* interaction terms, where *o* is the number of age-by-period interaction terms for that cohort. These tests indicate, generally speaking, whether the LFP of a cohort deviates significantly from the pattern defined by age and period main effects.

[Table 6 about here]

For example, for the 1950 cohort of black women, the *F* statistic of 8.028 ($p<0.001$) for Model 2a in Table 6 indicates that the group of age-by-period interactions for this cohort—i.e., that lie on the shaded diagonal corresponding to the 1950 cohort in Table 5—taken together, explains a significant portion of the variation in LFP. Generally speaking, among whites there seems to be a considerable amount of cohort deviation from age and period main effects, whereas fewer black cohorts deviate significantly from their general age and period pattern.

To quantify inter-cohort differences in terms of their deviations from the general age and period pattern, we computed the arithmetic mean of the group of age-by-period interactions for each cohort. The averages of the interaction terms and their p-values of the *t* tests in Step 3.1 are reported in the "Inter-Cohort" column in Table 7 for each race group. For each cohort, these *t*-tests test whether the average of the corresponding age-by-period interaction terms differ significantly from 0. Take again the 1950 cohort of black women for example: the average cohort deviation (0.077, $p<0.001$), estimated as the arithmetic mean of the group of age-by-period interaction terms in those shaded cells in Table 5, suggests that on average, the 1950 cohort had higher LFP than we would expect relative to their ages and periods. In general, the conclusion based on the average deviation statistics is consistent with the $F$ tests in Step 2; almost all white cohorts depart significantly from the general age and period main effects whereas only a few black cohorts appeared to. For whites, the baby boom cohorts born between 1950 and 1960 have significantly higher LFP rates than we would expect based on age and period main effects. For blacks, the 1950 cohort appears more likely and the 1935, 1980, and 1985 cohorts less likely to participate in the labor force relative to the general trend defined by the age and period main effects.

[Table 7 about here]

Did these average deviations in LFP among blacks and whites remain stable, decrease, or increase over the life course of those cohorts? According to the "cumulative advantage" hypothesis (Dannefer 1987, 2003; Hobcraft et al. 1982), cohorts with relatively high LFP rates should have progressively higher rates across the life course as they accumulate more experience, skills, and resources. In contrast, based on the "age-as-leveler" theory, we may expect cohorts with relatively low participation rates at young ages due to, e.g, being in school or having children, to catch up as they are older. That is, these theories and others discussed earlier suggest that LFP may change over the life course as a cohort ages.

We used the Step 3.2's $t$ test of a linear trend within a cohort's age-by-period interaction terms to examine the above hypotheses about life-course dynamics within cohort. The estimated slopes and the p-values are shown in the "Intra-Cohort" column in Table 7[11]. Each slope is estimated as the linear contrast in the age-by-period interaction terms contained in that cohort. Using the previous example of the 1950 cohort, these black women have an intra-cohort life-course slope of 0.05, estimated as the linear contrast in the age-by-period interaction terms in those shaded cells in Table 5. This positive slope is not statistically significant, suggesting that the 1950 cohort maintained (neither raised nor lowered) their relatively high LFP throughout their life course relative to their ages and times. The 1980 and 1985 cohorts also remained at a lower participation level relative to their ages and periods. Interestingly, the 1975 cohort had a negative slope (-0.107, $p<0.05$), although on average they did not deviate from the general age or period patterns. Combined with the age-by-period interactions in Table 5, it appears that although this cohort had an average LFP rate relative to their ages and periods, they had higher LFP rates in their 20s but average or lower participation when they were older.

---

[11] As noted earlier, we caution about the estimates for intra-cohort trends for the 1935 and 1990 cohorts. The effect estimates for these cohorts are determined by only two age-by-period

For whites, the 1950, 1955, and 1960 birth cohorts not only have higher LFP rates at young ages relative to the age and period main effects but they have (relatively) even higher participation rates at older ages: the intra-cohort life-course slopes for these cohorts are significantly positive and substantial in magnitude, suggesting that these cohorts' relative advantages amplified as they grew older. In contrast, the significant negative average deviations and life-course slopes for the 1965 through 1980 birth cohorts indicate that on average, members of those cohorts had a lower LFP relative to their age and period, and that this gap widened over the life course. Interestingly, for the 1945 birth cohort, their LFP rates seemed "compensatory";

---

interaction terms, so the linear trend in these effects may be different from the trend that would be observed if more interactions were available for these cohorts. The 1930 and 1995 cohorts have only one corresponding age-by-period interaction term, so no information about intra-cohort life-course change is available.

that is, this cohort's lower-than-expected LFP rates at younger ages were compensated by higher-than-expected rates at older ages so that their cohort average deviation is not statistically or substantively significant.

To what extent can we attribute the age, period, and cohort variations described above to changes in educational attainment and number of children? Specifically, given that education is often positively associated with LFP and the amount of formal schooling in the US has increased considerably in the last century (Fischer and Hout 2006), it would be interesting to see how changes in educational attainment have (differentially) affected LFP for blacks and whites. Similarly, since having children is negatively related to a woman's likelihood of participating in the labor force and the number of children per family in the US has been declining, we investigated how these changes may have been linked to temporal variation in women's LFP.

Our analytic strategy is to add—in separate analyses—measures for educational attainment and number of children. In each case, we ask how the age, period, and cohort variation in Models 1a and 2a are changed by holding constant the value of each factor. If we find, for example, that the age-by-period interaction is no longer present after adjusting for educational attainment, then we will conclude that the cohort patterns noted above are due to changes over time in educational attainment.

Models 1*b*, 1*c*, 2*b* and 2*c* in Table 4 report—Fig. 3b and 3c illustrate—how the age and period effects for white and black women's LFP were associated with changes in educational attainment and number of children, respectively. Controlling for educational attainment or child did not appear to affect general age trajectories except for the 20-24 age group, many of whom may have not completed their formal education. The shape of the period trends changed adjusting for educational attainment but not for number of children; had education not expanded, for a given education level, LFP would have gone down after the late 1990s among both blacks and whites. As shown in Table 7, neither the average deviations nor the life-course dynamics of the cohorts seem to change after considering education or children, although the higher-than-expected LFP rates of the early baby boom cohorts among whites seemed reduced.

*Conclusion*

In the example above, we examined age and period patterns and cohort deviations in women's LFP using the 1990-2017 CPS March Supplement data. The descriptive results of the age, period, and cohort patterns in LFP suggest that for blacks, the temporal variation in LFP can mostly be characterized by age and period trends, although a few cohorts deviated from these general trends. Whites had substantial cohort-related variation in their LFP that cannot be explained by age and period main effects. Most notably, the 1950-1960 cohorts of white women were more likely to participate in the labor force than expected, i.e., the actual LFP rates of these cohorts were higher than those described by the age groups and time periods that they experienced. Their higher-than-expected participation rates may be related to the lasting influences of the expansion of education and the feminist movement that they experienced during their youth.

Beyond average cohort deviations, we also found cumulative (dis)advantage over the life course in LFP among whites. That is, the higher-than-expected participation at young ages bred higher-than-expected participation at older ages among the 1950-1960 cohorts, while lower-than-expected participation at young ages led to lower-than-expected participation at older ages for other cohorts. On the one hand, this finding is consistent with the recent literature about the baby boom cohort's higher-than-expected LFP in older ages. Scholars posit that their relatively higher LFP may be attributed to their higher educational attainment and prior work experience (Goldin and Katz 2018; Goldin and Mitchell 2017). Our results show that their cumulatively higher LFP rates remained even after considering changes in education and number of children. Future research should address the hypothesis about the association between work experience and LFP for these cohorts.

On the other hand, the results suggest that white women born between 1965 through 1980 not only had a lower average LFP relative to their age and period, but even lower participation rates when they were older. This gap persists after adjusting for education and number of children. Although further research is required for a definite answer, the cumulative disadvantage of these cohorts casts doubt on optimism about a continuous growth of women's LFP.

The cumulatively higher-than-expected LFP rates of the 1950-60 cohorts of white women and cumulatively lower-than-expected rates of the younger cohorts also have important implications for understanding the effects of the Great Recession on women's LFP. Although the period main effects suggest no obvious decline in the late 2000s and the 2010s, this average trend may have masked the heterogeneous effects of the Great Recession: whereas the 1950-60 cohorts may not have been affected by the negative economic conditions, the adverse effect of the Great Recession was most obvious for the younger cohorts and may become even more pronounced when they are older. Future research may address this heterogeneity of period effects when the data are available.

**Discussion and Conclusion**

In this research, we discussed the gap between the conceptualization of cohort effects and their operationalization in traditional age-period-cohort (APC) models. We have developed a new APC method, called the APC-I model, for investigating inter-cohort deviations from age and period main effects and within-cohort life-course dynamics in social and demographic outcomes. Using Current Population Survey (CPS) data, we have shown how this model can be used to describe and explain age, period, inter-, and intra-cohort variation in black and white women's labor force participation (LFP).

The APC-I model has three advantages. First, like any two-way ANOVA model with an interaction, this model is identified, so it avoids the identification problem of the APC accounting model and allows inclusion of additional predictors. Second, the interpretation of the coefficient estimates of the APC-I model is meaningful and straightforward. Third, unlike traditional APC models that implicitly assume cohort effects are static through the life course, the APC-I model relaxes this assumption, allowing researchers to investigate life-course dynamics as a cohort ages.

Because the APC accounting framework has dominated the field for decades, it is not surprising if a reader tries to interpret the APC-I approach in terms of the accounting framework and concludes that the APC-I method ignores linear inter-cohort effects and only focuses on nonlinear effects defined in the accounting framework. Indeed, even among researchers who assume the APC accounting framework, a few methods have been proposed to focus on only the nonlinear cohort effects (Chauvel and Schröder 2015; Keyes et al. 2010; O'Brien et al. 2008). However, the APC-I model differs from such methods in two important ways: First, previous methods were all developed under the APC accounting framework's additive-cohort-effects assumption—that is, cohort effects are assumed to occur independently of social changes and the aging process *and* without differential effects of social changes for individuals of different ages. That is, previous methods do not question the validity of the APC accounting framework's assumption of additive cohort effects and thus offer little theoretical motivation for their operationalization of cohort effects. In contrast, we challenge this questionable and arbitrary assumption and have provided a detailed theoretical and conceptual justification for modeling cohort effects as a specific form of the age-by-period interaction. Second, although under certain circumstances cohort effects based on nonlinear cohort models may be numerically similar to the average cohort deviations estimated in Step 2 of the APC-I model, they have different meanings and interpretations and a nonlinear cohort effect in the APC accounting framework cannot be used to investigate life-course dynamics as in Step 3 of the APC-I model.

We stress that the APC-I model is not intended to recover the age, period, and cohort effects that the APC accounting model defines or to solve its identification problem. The cohort effects estimated in the APC-I model naturally differ from those in the accounting model because the additive and independent cohort effects operationalized in the APC accounting model depart from the interactive cohort effects modeled in the APC-I approach; the latter better represents, we argue, the sociological idea of what cohort effects are and when such effects can be observed. Interestingly, as we discussed earlier, Fienberg and Mason (1985) recognized the relationship between cohort effects and the age-by-period interaction, and they designed the APC accounting model with the explicit intent to model cohort effects as a particular form of the age-by-period interaction. In this sense, the APC-I model can be viewed a renewed effort to describe cohort effects as an age-by-period interaction.

As we discussed earlier, technically one could model age or period as the interaction of the other two effects, giving a model fit identical to the one given by the APC-I approach. The APC-I model makes an explicit and theoretically grounded choice to model cohorts using specific forms of the age-by-period interaction. However, the modeling choice ultimately depends on the relative theoretical importance of the three variables for specific research questions. For example, Harding and Jencks (2003) proposed an age-cohort model based on a parsimony criterion.[12] This model may be a viable alternative to the APC-I strategy with the presence of substantial qualitative or cross-over interactions, a circumstance in which the trend in the effect of period has a different direction depending on age and thus poses a challenge for interpreting main effects. More broadly, substantial qualitative age-by-period interactions may imply that a more sensible study design would compare life course trajectories between cohorts because each cohort has a distinct age or period pattern so that there is no general age or period trend. As Abbott (2005:7) argued, "we cannot write a history of periods. We customarily write the history of a population in terms of periods."

We used individual-level CPS data to demonstrate how to apply the APC-I model in empirical research. The APC-I model can also be used for aggregated data such as mortality, crime, or disease rates. However, for such aggregated data, the $(a-1) \cdot (p-1)$ freely-varying interaction terms are completely confounded with the error term because when arranged in the form of age-by-period cross-classification there is only one observation, i.e., no replication, per cell (e.g., mortality rate for age 80 in 2000). If the outcome is a binary or count variable analyzed using logistic or Poisson regression, one can test the interaction even though the model including the interaction terms is saturated. For continuous data modeled as normal, this no-replication is indeed a problem. For such data, it is still possible to perform the global and local deviance tests in Steps 1 and 2 using various statistical solutions that involve testing simpler, less-than-saturated models (e.g., Tukey's test of additivity [Tukey 1949]). Of course, such a test is only a partial solution to the no-replication problem, but it gives researchers some ability to investigate whether cohort effects exist by detecting departures from additive effects of age and periods.

---

[12] Adding age-cohort interactions to an age-cohort model may be difficult for repeated cross-sectional designs like the General Social Survey and the CPS because older and younger cohorts are often observed only one or two times, limiting the ability to estimate age-period interactions.

In addition to describing "raw" trends, the APC-I model can be used to investigate temporal trends in the outcome by modeling the ways in which possible explanatory factors affect these trends. To this extent, this method echoes the ideas behind the proxy variable approach (Heckman and Robb 1985), the age-period-cohort characteristics model (O'Brien 2000), the mechanism-based model (Winship and Harding 2008), and a bounding strategy (Fosse and Winship 2018), which specify the theoretical mechanisms through which age, period, and/or cohort affects the outcome. The importance of theoretical thinking in informing model specification and interpretation cannot be overstressed. In their insightful article, Fienberg and Mason (1985) encouraged researchers to "begin with conceptualization and attempt to move toward explicit measurement, in order to test understanding of the interaction [i.e., the cohort effects]" (p. 83). To the extent that the APC-I model is explicitly tied to the conceptualization of cohort effects in sociological and demographic literatures, we believe that the APC-I model has promise for advancing APC research.

Another potential use of the APC-I model is that it allows constructing, estimating, and testing more theoretically meaningful cohorts. Many APC studies, including our analysis of women's LFP, use "convenient" birth cohorts, cohorts with memberships that are not constructed based on a theoretical account or prior knowledge about the distinctive experience of a group of people but rather are determined by age and period intervals. Such cohorts might not experience distinctive social changes during their critical ages, so they may not be considered a cohort in a substantive sense. With data that have finer, e.g., one-year or two-year, age and period intervals, a researcher may construct more meaningful cohorts by drawing cohort boundaries based on prior knowledge about a cohort's distinctive experience. The effects of such cohorts are thus represented by age-by-period interaction terms that lie on *multiple* diagonals in the age-by-period table. The same three-step procedure we described for the APC-I model can be applied to examine the age-by-period interaction terms that lie on those multiple diagonals.

Lastly, although the APC-I model was designed for APC analysis, the conceptual critique and methodological ideas can be extended to many other fields in which focal explanatory variables are exactly related. For example, scholars of status inconsistency study the likelihood of a person attaining higher or lower socioeconomic status than their parents and the consequences of changes in status for various outcomes including happiness, marriage, and health conditions. Also, researchers of assortative mating are interested in how marriage forms between persons of the same or different levels of educational attainment, and the implications of such educational homogeneity or heterogeneity for marriage duration, life satisfaction, and other aspects of economic and health well-being. Despite long-standing interest in these areas among sociologists and demographers, these lines of scholarship suffer from effectively the same methodological problem as APC analysis: the third variable is determined by the other two. Specifically, in status inconsistency studies, status inconsistency equals adult socioeconomic status minus status of their parents; in educational homogamy research, educational difference equals husband's education minus wife's education. Several methods have been developed to address this estimation problem, but none is satisfactory from a statistical point of view (Hope 1975; Houle 2011; Sobel 1981, 1985). The APC-I model developed in this paper can potentially be modified to address these important sociological issues.

**Table 1. Unobserved Parameters in Models (1) and (5).**

|  |  |  | Period |  |  |  |  |
|---|---|---|---|---|---|---|---|
|  |  |  | *1* | *2* | *3* | *4* | *5* |
| **Parameters in Model (1)** | Age | 1 | $\mu+\alpha_1+\beta_1+\gamma_5$ | $\mu+\alpha_1+\beta_2+\gamma_6$ | $\mu+\alpha_1+\beta_3+\gamma_7$ | $\mu+\alpha_1+\beta_4+\gamma_8$ | $\mu+\alpha_1+\beta_5+\gamma_9$ |
|  |  | 2 | $\mu+\alpha_2+\beta_1+\gamma_4$ | $\mu+\alpha_2+\beta_2+\gamma_5$ | $\mu+\alpha_2+\beta_3+\gamma_6$ | $\mu+\alpha_2+\beta_4+\gamma_7$ | $\mu+\alpha_2+\beta_5+\gamma_8$ |
|  |  | 3 | $\mu+\alpha_3+\beta_1+\gamma_3$ | $\mu+\alpha_3+\beta_2+\gamma_4$ | $\mu+\alpha_3+\beta_3+\gamma_5$ | $\mu+\alpha_3+\beta_4+\gamma_6$ | $\mu+\alpha_3+\beta_5+\gamma_7$ |
|  |  | 4 | $\mu+\alpha_4+\beta_1+\gamma_2$ | $\mu+\alpha_4+\beta_2+\gamma_3$ | $\mu+\alpha_4+\beta_3+\gamma_4$ | $\mu+\alpha_4+\beta_4+\gamma_5$ | $\mu+\alpha_4+\beta_5+\gamma_6$ |
|  |  | 5 | $\mu+\alpha_5+\beta_1+\gamma_1$ | $\mu+\alpha_5+\beta_2+\gamma_2$ | $\mu+\alpha_5+\beta_3+\gamma_3$ | $\mu+\alpha_5+\beta_4+\gamma_4$ | $\mu+\alpha_5+\beta_5+\gamma_5$ |
| **Parameters in Model (5)** | Age | 1 | $\mu+\alpha_1+\beta_1+\alpha\beta_{11(5)}$ | $\mu+\alpha_1+\beta_2+\alpha\beta_{12(6)}$ | $\mu+\alpha_1+\beta_3+\alpha\beta_{13(7)}$ | $\mu+\alpha_1+\beta_4+\alpha\beta_{14(8)}$ | $\mu+\alpha_1+\beta_5+\alpha\beta_{15(9)}$ |
|  |  | 2 | $\mu+\alpha_2+\beta_1+\alpha\beta_{21(4)}$ | $\mu+\alpha_2+\beta_2+\alpha\beta_{22(5)}$ | $\mu+\alpha_2+\beta_3+\alpha\beta_{23(6)}$ | $\mu+\alpha_2+\beta_4+\alpha\beta_{24(7)}$ | $\mu+\alpha_2+\beta_5+\alpha\beta_{25(8)}$ |
|  |  | 3 | $\mu+\alpha_3+\beta_1+\alpha\beta_{31(3)}$ | $\mu+\alpha_3+\beta_2+\alpha\beta_{32(4)}$ | $\mu+\alpha_3+\beta_3+\alpha\beta_{33(5)}$ | $\mu+\alpha_3+\beta_4+\alpha\beta_{34(6)}$ | $\mu+\alpha_3+\beta_5+\alpha\beta_{35(7)}$ |
|  |  | 4 | $\mu+\alpha_4+\beta_1+\alpha\beta_{41(2)}$ | $\mu+\alpha_4+\beta_2+\alpha\beta_{42(3)}$ | $\mu+\alpha_4+\beta_3+\alpha\beta_{43(4)}$ | $\mu+\alpha_4+\beta_4+\alpha\beta_{44(5)}$ | $\mu+\alpha_4+\beta_5+\alpha\beta_{45(6)}$ |
|  |  | 5 | $\mu+\alpha_5+\beta_1+\alpha\beta_{51(1)}$ | $\mu+\alpha_5+\beta_2+\alpha\beta_{52(2)}$ | $\mu+\alpha_5+\beta_3+\alpha\beta_{53(3)}$ | $\mu+\alpha_5+\beta_4+\alpha\beta_{54(4)}$ | $\mu+\alpha_5+\beta_5+\alpha\beta_{55(5)}$ |

Note: $\alpha_i$ denotes the mean difference from the global mean $\mu$ associated with the ith age category; $\beta_j$ denotes the mean difference from $\mu$ associated with the jth period; $\gamma_k$ denotes the mean difference from $\mu$ due to the membership in the kth cohort in Model (1); $\alpha\beta_{ij(k)}$ denotes the mean difference from age main effects $\alpha_i$ and period main effects $\beta_j$ associated with ijth age-by-period interaction in Model (5).

**Table 2. Testing Theories about Intra-Cohort Life-Course Dynamics Using the Three-Step Procedure.**

|  |  | Sign of Intra-Cohort Life-Course Dynamics (Step 3.2) | | |
|---|---|---|---|---|
|  |  | + | 0 | - |
| **Sign of Average Cohort Deviation (Step 3.1)** | + | cumulative advantage | constant | leveling |
|  | 0 | leveling | no clear pattern | leveling |
|  | - | leveling | constant | cumulative disadvantage |

Note: + denotes significant and positive estimate; 0 denotes non-significant estimates; - denotes significant and negative estimates.

**Table 3. Descriptive Statistics for All Analytic Variables in the Current Population Survey Data, 1990-2017.**

| | White Women | | | | | Black Women | | | | |
|---|---|---|---|---|---|---|---|---|---|---|
| Description | N | Mean | S.D. | Min. | Max. | N | Mean | S.D. | Min. | Max. |
| Labor Force Participation (LABFORCE; 1=in the labor force; 0=no in the labor force) | 1,213,497 | 0.72 | (0.45) | 0 | 1 | 181,064 | 0.71 | (0.45) | 0 | 1 |
| Age at time of survey (AGE) | 1,213,497 | 40.83 | (12.00) | 20 | 64 | 181,064 | 40.57 | (12.38) | 20 | 64 |
| Survey year (YEAR) | 1,213,497 | - | - | 1990 | 2017 | 181,064 | - | - | 1990 | 2017 |
| Birth year (YEAR - AGE) | 1,213,497 | - | - | 1930 | 1995 | 181,064 | - | - | 1930 | 1995 |
| Educational attainment (EDUC; 3=college or more; 2=some college; 1=high school; 0=less than high school) | 1,213,497 | 1.74 | (0.99) | 0 | 3 | 181,064 | 1.55 | (0.96) | 0 | 3 |
| Number of Children (NCHILD; 3=3 children or more; 2=2 children; 1=1 child; 0=no children) | 1,213,497 | 1.07 | (1.08) | 0 | 3 | 181,064 | 1.01 | (1.07) | 0 | 3 |

Note: Analysis includes women respondents who participated in the 1990 through 2017 CPS surveys and for whom labor force participation, year of birth, gender, and race are available. Words in all caps are CPS variable names.

**Table 4. Estimated Age and Period Main Effects on Labor Force Participation, with and without Adjustment for Education, CPS 1990-2017.**

|  |  | White | | | Black | | |
|---|---|---|---|---|---|---|---|
|  |  | Model 1a | Model 1b | Model 1c | Model 2a | Model 2b | Model 2c |
| **Intercept** |  | 0.899 *** | 0.782 *** | 0.732 *** | 0.863 *** | 0.886 *** | 0.827 *** |
| **Education** | <H.S. | — | -0.892 *** | — | — | -0.945 *** | — |
|  | H.S. | — | -0.011 *** | — | — | -0.115 *** | — |
|  | Some Col. | — | 0.253 *** | — | — | 0.225 *** | — |
|  | ≥B.A. | — | 0.649 *** | — | — | 0.835 *** | — |
| **Number of Children** | 0 | — | — | 0.417 *** | — | — | 0.034 ** |
|  | 1 | — | — | 0.193 *** | — | — | 0.183 *** |
|  | 2 | — | — | -0.045 *** | — | — | 0.134 *** |
|  | ≥3 | — | — | -0.566 *** | — | — | -0.351 *** |
| **Age** | 20-24 | -0.011 | 0.017 * | -0.163 *** | -0.185 *** | -0.162 *** | -0.202 *** |
|  | 25-29 | 0.228 *** | 0.198 *** | 0.225 *** | 0.288 *** | 0.246 *** | 0.304 *** |
|  | 30-34 | 0.151 *** | 0.120 *** | 0.293 *** | 0.370 *** | 0.316 *** | 0.416 *** |
|  | 35-39 | 0.199 *** | 0.175 *** | 0.409 *** | 0.492 *** | 0.435 *** | 0.546 *** |
|  | 40-44 | 0.320 *** | 0.300 *** | 0.490 *** | 0.425 *** | 0.382 *** | 0.441 *** |
|  | 45-49 | 0.340 *** | 0.334 *** | 0.397 *** | 0.277 *** | 0.270 *** | 0.260 *** |
|  | 50-54 | 0.131 *** | 0.146 *** | 0.060 *** | 0.003 | 0.030 | -0.028 |
|  | 55-59 | -0.289 *** | -0.258 *** | -0.445 *** | -0.466 *** | -0.395 *** | -0.500 *** |
|  | 60-64 | -1.068 *** | -1.032 *** | -1.266 *** | -1.205 *** | -1.122 *** | -1.237 *** |
| **Period** | 1990-94 | -0.097 *** | -0.003 | -0.108 *** | -0.203 *** | -0.041 ** | -0.201 *** |
|  | 1995-99 | 0.012 * | 0.059 *** | 0.008 | -0.019 | 0.063 *** | -0.018 |
|  | 2000-04 | 0.046 *** | 0.057 *** | 0.043 *** | 0.093 *** | 0.115 *** | 0.090 *** |
|  | 2005-09 | 0.024 *** | 0.001 | 0.027 *** | 0.068 *** | 0.031 * | 0.068 *** |
|  | 2010-14 | 0.011 * | -0.045 *** | 0.018 ** | 0.026 | -0.068 *** | 0.027 * |
|  | 2015-17 | 0.005 | -0.071 *** | 0.011 | 0.035 * | -0.100 *** | 0.034 * |
| **Cohort** |  | | | (See Table 7) | | | |
| **N** |  | | 1,213,497 | | | 181,064 | |

Note: Analysis includes women respondents aged 20 to 64 who participated in the 1990 through 2017 CPS surveys in years for whom labor force participation status, year of birth, educational attainment, and number of children are available. Table figures represent weighted logistic APC-I model estimates using the sum-to-zero coding. ***=p<0.001 ; ** = p < 0.01 ; * = p < 0.05.

**Table 5. Estimates of Age-by-Period Interaction Terms in Models 1a and 2a.**

**White Women: Model 1a**

|  |  | Period |  |  |  |  |  |
|---|---|---|---|---|---|---|---|
|  |  | 1990-94 | 1995-99 | 2000-04 | 2005-09 | 2010-14 | 2015-17 |
|  | 20-24 | 0.152 *** | 0.097 *** | 0.040 ** | -0.050 *** | -0.129 *** | -0.109 *** |
|  | 25-29 | 0.070 *** | 0.061 *** | -0.026 | -0.060 *** | -0.082 *** | 0.038 * |
|  | 30-34 | 0.065 *** | 0.029 | -0.009 | -0.044 ** | -0.023 | -0.018 |
|  | 35-39 | 0.130 *** | 0.071 *** | -0.024 | -0.085 *** | -0.033 * | -0.059 *** |
| Age | 40-44 | 0.167 *** | 0.085 *** | 0.031 * | -0.032 * | -0.101 *** | -0.149 *** |
|  | 45-49 | 0.032 * | 0.074 *** | 0.074 *** | 0.005 | -0.085 *** | -0.099 *** |
|  | 50-54 | -0.118 *** | -0.014 | 0.069 *** | 0.061 *** | 0.041 ** | -0.040 * |
|  | 55-59 | -0.224 *** | -0.151 *** | -0.057 *** | 0.111 *** | 0.158 *** | 0.164 *** |
|  | 60-64 | -0.274 *** | -0.251 *** | -0.097 *** | 0.095 *** | 0.254 *** | 0.272 *** |

**Black Women: Model 2a**

|  |  | Period |  |  |  |  |  |
|---|---|---|---|---|---|---|---|
|  |  | 1990-94 | 1995-99 | 2000-04 | 2005-09 | 2010-14 | 2015-17 |
|  | 20-24 | 0.000 | 0.109 ** | -0.012 | -0.107 ** | 0.019 | -0.009 |
|  | 25-29 | -0.108 ** | 0.123 ** | 0.039 | 0.015 | -0.085 * | 0.015 |
|  | 30-34 | -0.004 | 0.014 | 0.081 * | 0.059 | -0.134 *** | -0.016 |
|  | 35-39 | 0.051 | 0.018 | 0.062 | -0.038 | -0.042 | -0.051 |
| Age | 40-44 | 0.161 *** | -0.062 | -0.020 | -0.042 | -0.018 | -0.019 |
|  | 45-49 | 0.033 | -0.054 | -0.031 | 0.036 | -0.017 | 0.034 |
|  | 50-54 | 0.019 | 0.045 | 0.022 | -0.098 ** | 0.011 | 0.001 |
|  | 55-59 | -0.146 ** | -0.030 | -0.023 | 0.084 * | 0.096 ** | 0.019 |
|  | 60-64 | -0.005 | -0.163 ** | -0.118 ** | 0.092 * | 0.171 *** | 0.024 |

Note: Table figures represent estimated age-by-period interaction terms coded to sum to zero in weighted logistic APC-I Models 1a for whites and Model 2a for blacks. Models 1a and 2a modeled white and black women's labor force partication using age, period, and their interactions without covariates. ***=p<0.001 ; ** = p < 0.01 ; * = p < 0.05.

**Table 6. Local Deviance Tests for Cohort Devitions in White and Black Women's Labor Force Participation, with and without Adjusting for Education and Number of Children, CPS 1990-2017.**

|  |  | Whites | | | | | | | | |
|---|---|---|---|---|---|---|---|---|---|---|
|  |  | Model 1a | | | Model 1b | | | Model 1c | | |
|  |  | Statistic | df1 | df2 | Statistic | df1 | df2 | Statistic | df1 | df2 |
| Cohort | 1930 | 351.981 | 1 | 1213482 *** | 193.791 | 1 | 1213479 *** | 314.183 | 1 | 1213479 *** |
|  | 1935 | 235.864 | 2 | 1213481 *** | 125.406 | 2 | 1213478 *** | 218.123 | 2 | 1213478 *** |
|  | 1940 | 74.146 | 3 | 1213480 *** | 40.610 | 3 | 1213477 *** | 79.654 | 3 | 1213477 *** |
|  | 1945 | 11.257 | 4 | 1213479 *** | 9.280 | 4 | 1213476 *** | 13.950 | 4 | 1213476 *** |
|  | 1950 | 75.747 | 5 | 1213478 *** | 39.849 | 5 | 1213475 *** | 63.536 | 5 | 1213475 *** |
|  | 1955 | 58.460 | 6 | 1213477 *** | 32.930 | 6 | 1213474 *** | 54.705 | 6 | 1213474 *** |
|  | 1960 | 15.425 | 6 | 1213477 *** | 13.376 | 6 | 1213474 *** | 19.802 | 6 | 1213474 *** |
|  | 1965 | 12.758 | 6 | 1213477 *** | 7.579 | 6 | 1213474 *** | 11.945 | 6 | 1213474 *** |
|  | 1970 | 40.432 | 6 | 1213477 *** | 26.409 | 6 | 1213474 *** | 41.044 | 6 | 1213474 *** |
|  | 1975 | 23.826 | 5 | 1213478 *** | 16.199 | 5 | 1213475 *** | 23.705 | 5 | 1213475 *** |
|  | 1980 | 9.642 | 4 | 1213479 *** | 7.876 | 4 | 1213476 *** | 11.875 | 4 | 1213476 *** |
|  | 1985 | 10.817 | 3 | 1213480 *** | 7.487 | 3 | 1213477 *** | 17.709 | 3 | 1213477 *** |
|  | 1990 | 37.930 | 2 | 1213481 *** | 33.320 | 2 | 1213478 *** | 48.604 | 2 | 1213478 *** |
|  | 1995 | 30.435 | 1 | 1213482 *** | 27.156 | 1 | 1213479 *** | 52.249 | 1 | 1213479 *** |

|  |  | Blacks | | | | | | | | |
|---|---|---|---|---|---|---|---|---|---|---|
|  |  | Model 2a | | | Model 2b | | | Model 2c | | |
|  |  | Statistic | df1 | df2 | Statistic | df1 | df2 | Statistic | df1 | df2 |
| Cohort | 1930 | 0.419 | 1 | 181049 | 3.523 | 1 | 181046 | 0.528 | 1 | 181046 |
|  | 1935 | 12.171 | 2 | 181048 *** | 2.547 | 2 | 181045 | 12.740 | 2 | 181045 *** |
|  | 1940 | 4.907 | 3 | 181047 ** | 4.449 | 3 | 181044 ** | 5.095 | 3 | 181044 ** |
|  | 1945 | 1.462 | 4 | 181046 | 1.330 | 4 | 181043 | 1.663 | 4 | 181043 |
|  | 1950 | 8.028 | 5 | 181045 *** | 3.750 | 5 | 181042 ** | 8.077 | 5 | 181042 *** |
|  | 1955 | 3.212 | 6 | 181044 ** | 3.802 | 6 | 181041 *** | 3.245 | 6 | 181041 ** |
|  | 1960 | 0.439 | 6 | 181044 | 0.509 | 6 | 181041 | 0.393 | 6 | 181041 |
|  | 1965 | 1.891 | 6 | 181044 | 2.390 | 6 | 181041 * | 1.792 | 6 | 181041 |
|  | 1970 | 2.420 | 6 | 181044 * | 2.172 | 6 | 181041 * | 2.769 | 6 | 181041 * |
|  | 1975 | 2.651 | 5 | 181045 * | 1.951 | 5 | 181042 | 3.049 | 5 | 181042 ** |
|  | 1980 | 3.406 | 4 | 181046 ** | 2.755 | 4 | 181043 * | 3.252 | 4 | 181043 * |
|  | 1985 | 4.545 | 3 | 181047 ** | 2.810 | 3 | 181044 * | 5.409 | 3 | 181044 ** |
|  | 1990 | 0.332 | 2 | 181048 | 1.562 | 2 | 181045 | 0.152 | 2 | 181045 |
|  | 1995 | 0.023 | 1 | 181049 | 0.682 | 1 | 181046 | 0.032 | 1 | 181046 |

Note: Table figures represent generalized *F* statistics and their associated degrees of freedom of the local deviance tests for each cohort. Models 1a and 2a does not include covariates; Models 1b and 2b include edcuational attainment as a covariate; Models 1c and 2c include number of children as a covariate. ***=p<0.001 ; ** = p < 0.01 ; * = p < 0.05.

**Table 7. Estimated Inter-Cohort Differences and Intra-Cohort Dynamics in White and Black Women's Labor Force Participation, with and without Adjusting for Education and Number of Children, CPS 1990-2017.**

| | | Whites | | | | | |
|---|---|---|---|---|---|---|---|
| | | Model 1a | | Model 1b | | Model 1c | |
| | | Inter-Cohort | Intra-Cohort | Inter-Cohort | Intra-Cohort | Inter-Cohort | Intra-Cohort |
| Cohort | 1930 | -0.274 *** | NA | -0.211 *** | NA | -0.259 *** | NA |
| | 1935 | -0.237 *** | -0.019 | -0.179 *** | -0.019 | -0.230 *** | -0.024 |
| | 1940 | -0.122 *** | 0.015 | -0.093 *** | -0.002 | -0.130 *** | 0.019 |
| | 1945 | 0.014 | 0.033 * | 0.003 | 0.012 | -0.004 | 0.040 * |
| | 1950 | 0.135 *** | 0.067 *** | 0.096 *** | 0.055 *** | 0.119 *** | 0.085 *** |
| | 1955 | 0.130 *** | 0.110 *** | 0.099 *** | 0.104 *** | 0.127 *** | 0.125 *** |
| | 1960 | 0.063 *** | 0.045 ** | 0.061 *** | 0.055 *** | 0.071 *** | 0.056 *** |
| | 1965 | -0.014 * | -0.107 *** | -0.010 | -0.085 *** | 0.003 | -0.100 *** |
| | 1970 | -0.014 * | -0.217 *** | -0.015 * | -0.173 *** | 0.000 | -0.209 *** |
| | 1975 | -0.031 *** | -0.158 *** | -0.021 ** | -0.136 *** | -0.018 * | -0.166 *** |
| | 1980 | -0.025 ** | -0.058 *** | -0.006 | -0.061 *** | -0.022 ** | -0.074 *** |
| | 1985 | -0.050 *** | 0.023 | -0.042 *** | 0.015 | -0.065 *** | 0.006 |
| | 1990 | -0.046 *** | 0.119 *** | -0.047 *** | 0.110 *** | -0.078 *** | 0.102 *** |
| | 1995 | -0.109 *** | NA | -0.105 *** | NA | -0.143 *** | NA |

| | | Blacks | | | | | |
|---|---|---|---|---|---|---|---|
| | | Model 2a | | Model 2b | | Model 2c | |
| | | Inter-Cohort | Intra-Cohort | Inter-Cohort | Intra-Cohort | Inter-Cohort | Intra-Cohort |
| Cohort | 1930 | -0.005 | NA | 0.108 * | NA | -0.008 | NA |
| | 1935 | -0.155 *** | -0.012 | -0.083 * | -0.021 | -0.158 *** | -0.015 |
| | 1940 | -0.043 | -0.097 * | -0.011 | -0.135 ** | -0.045 | -0.098 * |
| | 1945 | 0.037 | 0.025 | 0.037 | -0.003 | 0.040 | 0.028 |
| | 1950 | 0.077 *** | 0.050 | 0.034 | 0.016 | 0.074 *** | 0.055 |
| | 1955 | -0.003 | 0.033 | -0.038 * | 0.012 | -0.006 | 0.039 |
| | 1960 | 0.010 | 0.018 | -0.004 | 0.019 | 0.009 | 0.020 |
| | 1965 | -0.015 | 0.042 | -0.013 | 0.063 | -0.013 | 0.044 |
| | 1970 | 0.030 | -0.045 | 0.010 | 0.029 | 0.037 * | -0.046 |
| | 1975 | 0.029 | -0.107 * | 0.033 | -0.069 | 0.035 | -0.112 ** |
| | 1980 | -0.045 * | -0.060 | -0.026 | -0.066 | -0.042 * | -0.050 |
| | 1985 | -0.069 ** | 0.065 | -0.049 * | 0.063 | -0.081 *** | 0.059 |
| | 1990 | 0.017 | -0.002 | 0.048 | -0.003 | 0.006 | -0.013 |
| | 1995 | -0.009 | NA | 0.039 | NA | -0.011 | NA |

Note: Table figures in "Inter-Cohort" columns are avearaged age-by-period interaction estimates corresponding to each cohort in the weighted logistic APC-I models using the sum-to-zero coding. Table figures in "Intra-Cohort" columns are estimated linear slopes in the age-by-period interaction estimates contained in each cohort in the weighted logistic APC-I models using the sum-to-zero coding. Models 1a and 2a does not include covariates; Models 1b and 2b include ecuational attainment as a covariate; Models 1c and 2c include number of children as a covariate. ***=p<0.001 ; ** = p < 0.01 ; * = p < 0.05.

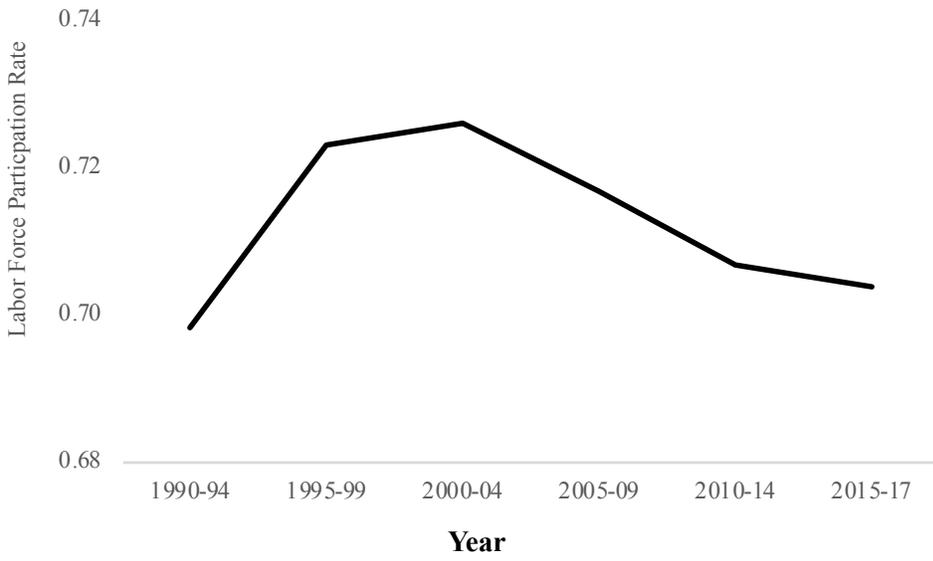

Figure 1. Labor Force Participation Rates among Women Aged 20 to 64 in the US, CPS 1990-2017.

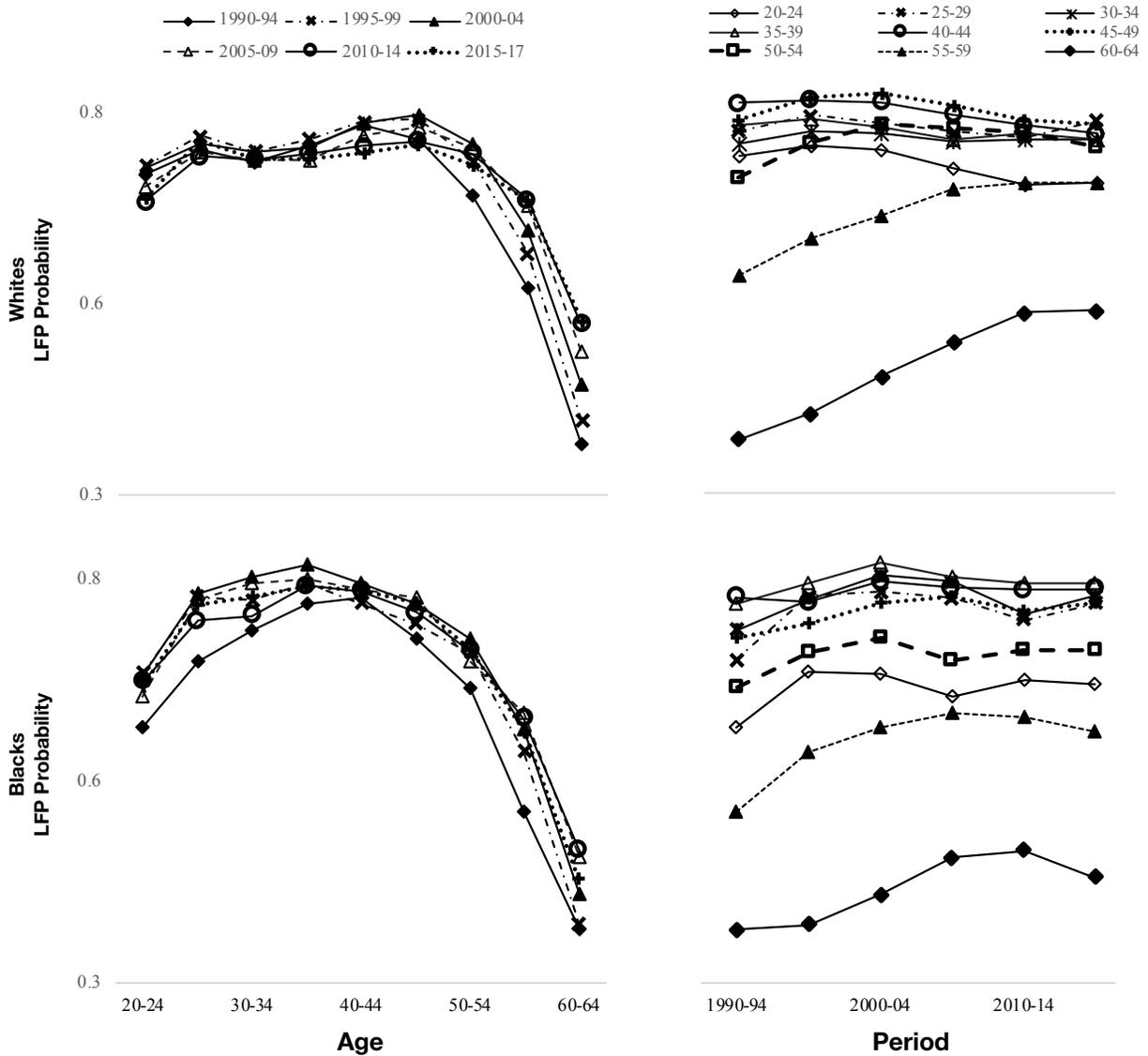

Figure 2. Period-Specific Age Patterns and Age-Specific Period Patterns in White and Black Women's Labor Force Participation Probability, CPS 1990-2017.

Note: Figures represent estimated LFP probabilities converted from the coefficient estimates of the weighted logistic APC-I models 1a for whites and 2a for blacks in Table 4. Period-specific age patterns are based on the estimated intercept, age main effects, and the age-period interaction terms for each time period from Models 1a for whites and 2a for blacks. Age-specific period patterns are based on the estimated intercept, period main effects, and the age-period interaction terms for each age group from Models 1a for whites and 2a for blacks.

**Figure 3a. No Covariates**

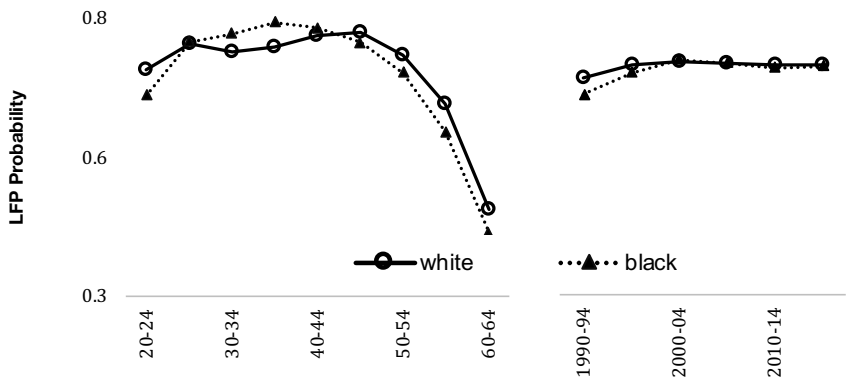

**Figure 3b. Adjusting for Education**

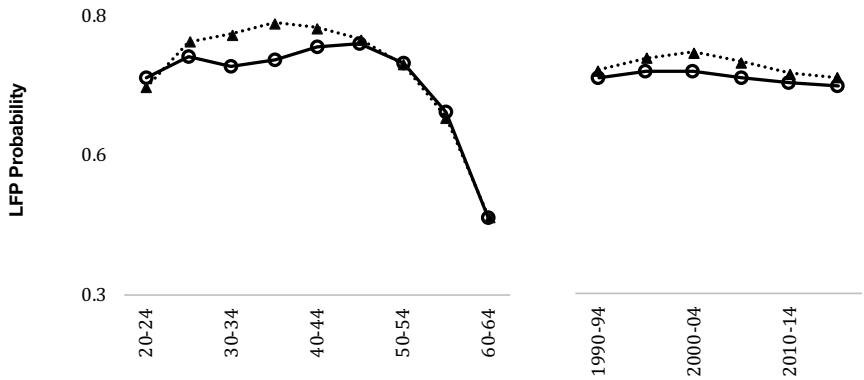

**Figure 3c. Adjusting for Number of Children**

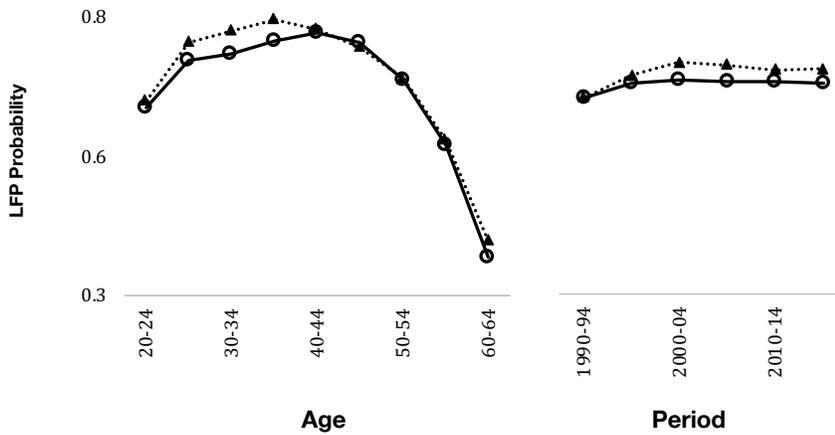

Figure 3. Estimated Age and Period Main Effects in White and Black Women's Labor Force Participation, with and without Adjusting for Education and Number of Children, CPS 1990-2017.

Note: Figures represent estimated LFP probabilities converted from the coefficient estimates of the weighted logistic APC-I models 1a, 1b, and 1c for whites and 2a, 2b, and 2c for blacks, respectively. Age and period patterns are based on the estimated intercepts and age or period main effects in Table 4.